\documentstyle[epsfig,rotating,12pt,english]{article}
\newcommand{\alsmz}{\mathrm{\alpha_s(\MZt)}}
\newcommand{\als}{\mathrm{\alpha_s}}
\newcommand{\MZ}{\mathrm{M_Z}}
\newcommand{\GZ}{\mathrm{\Gamma_Z}}
\newcommand{\MZt}{\mathrm{M_Z^2}}
\newcommand{\GZt}{\mathrm{\Gamma_Z^2}}
\newcommand{\sigh}{\mathrm{\sigma^0_{had}}}
\newcommand{\sigl}{\mathrm{\sigma^0_l}}
\newcommand{\Rl}{\mathrm{R}_\ell}
\newcommand{\Rb}{\mathrm{R^0_b}}
\newcommand{\Gh}{\mathrm{\Gamma_{had}}}
\newcommand{\Gb}{\mathrm{\Gamma_b}}
\newcommand{\Ginv}{\mathrm{\Gamma_{inv}}}
\newcommand{\Ge}{\mathrm{\Gamma_e}}
\newcommand{\Gf}{\mathrm{\Gamma_f}}
\newcommand{\Gm}{\mathrm{\Gamma_\mu}}
\newcommand{\Gt}{\mathrm{\Gamma_\tau}}

\newcommand{\Gl}{\mathrm{\Gamma_l}}

\newcommand{\alem}{\mathrm{\alpha(\MZt)}}
\newcommand{\MH}{\mathrm{M_H}}
\newcommand{\Mt}{\mathrm{M_t}}
\newcommand{\MW}{\mathrm{M_W}}
\newcommand{\ee}{\mathrm{e^+e^-}}

\newcommand{\qq}{\mathrm{q\bar{q}}}
\newcommand{\ff}{\mathrm{f\bar{f}}}

\newcommand{\eeee}{\ee\rightarrow \ee}

\newcommand{\eeff}{\ee\rightarrow \ff}
\newcommand{\sineff}{\mathrm{\sin^2\theta^{lept}_{eff}}}
\newcommand{\sinW}{\mathrm{\sin^2\theta_W}}
\begin{document}
\begin{titlepage}
\pagestyle{empty}
\vskip 2.0 cm
\begin{flushright}
LAL 98-67\\
31st August 1998
\end{flushright}
\vskip 1.0 cm
\begin{center}{
{\LARGE\bf Extraction of $\als$ and constraint on the Higgs mass from
Electroweak fits at the Z resonance}
\vskip 1.0 cm
 Edwige TOURNEFIER\footnote{e-mail: Edwige.Tournefier@cern.ch}\\
Laboratoire de l'Acc\'el\'erateur Lin\'eaire\\
 IN2P3-CNRS et Universit\'e de Paris-Sud,\\ BP34,
 F-91898 Orsay Cedex}
\end{center}
\vskip 1.5 cm

\begin{abstract}
The determination of the Z lineshape parameters at LEP1 is presented
and the value of $\alsmz$ is derived from these measurements.
The constraint on the Higgs mass obtained from a global fit to LEP1 and SLC 
data is also given.
\end{abstract}
\vskip 3cm
\begin{center}
{\it
Talk given at the 10th International Seminar\\
``QUARKS' 98'' \\
Suzdal, Russia, May 18-24, 1998}
\end{center}
\end{titlepage}
\newpage
\pagestyle{plain}
%
%
\section{Introduction}
\label{sec:intro}
From 1990 to October 1995 the LEP $\ee$ storage ring was operated
at center-of-mass energies close to the Z mass (called LEP1 program).
LEP1 data have been collected and analysed by the 4 LEP experiments
ALEPH\cite{aleph}, DELPHI\cite{delphi}, L3\cite{l3} and OPAL\cite{opal},
the electroweak results and their combination are given in
 Ref.\cite{lepew98} and are still preliminary.
Since 1992 the SLD detector is taking data on the SLC $\ee$ storage ring
operating at center-of-mass energies also close to  
$\MZ$ with polarised electron beam.
The preliminary results from SLD include 1992 to 1997 data \cite{SLD}.
\\
The cross sections and the asymmetries of the reactions $\eeff(\gamma)$ 
measured at LEP and SLC are sensitive, through radiative corrections,
to the following Standard Model parameters: the strong coupling 
constant $\als$, the Higgs mass $\MH$ and the top mass $\Mt$. 
The value of $\als$ is mainly determined by the LEP total cross section 
measurements,
while the asymmetries measured at LEP and SLC are most sensitive to $\MH$.
We will concentrate on the determination of $\als$ and therefore on the
determination of the Z lineshape parameters at LEP. 
%
%
\section{The Z lineshape and the fitting procedure}
\label{sec:Zline}
At LEP, an integrated luminosity of 110 $\mathrm{pb^{-1}}$ per experiment
has been
accumulated at the Z peak and about 40 $\mathrm{pb^{-1}}$ off peak, mostly
at $\MZ \pm1.8$~GeV. About 4 million hadronic and 0.5 million leptonic
Z decays have been collected by each of the four LEP experiments.
This large sample allows a precise determination of the Z boson properties:
the Z mass $\MZ$, the Z width $\GZ$,
the total hadronic cross section at the pole $\sigh$ and the ratio of 
hadronic to leptonic pole cross sections $\Rl=\sigh/\sigl\equiv\Gh/\Gl$.
\\
These parameters have the advantage of being almost uncorrelated:
\begin{itemize}
\item $\MZ$ is determined by the position of the peak and therefore 
depends on the absolute energy scale. Its uncertainty is dominated 
by the uncertainty on the LEP energy which is about 1.5~MeV\cite{lepew98}. 
\item $\GZ$ is determined by the width of the Z resonance
 and therefore by peak and off peak relative cross section measurements.
The uncertainty on $\GZ$
comes mainly from uncorrelated errors on the off peak energy measurement 
($\sim$1.5~MeV\cite{lepew98}) and from off peak statistics.
\item $\sigh$ is determined by the height of the resonance and is derived from
the measurement of the hadronic cross section.
The statistical and systematic uncertainties of the selections are 
typically of the order of $0.5\times 10^{-3}$ and $0.8\times 10^{-3}$
respectively, for each experiment\cite{lepew98}. The error on $\sigh$
is dominated by
the theoretical uncertainty on the luminosity which is $1.1\times 10^{-3}$.
Note that a recent study on the theoretical precision of the LEP 
luminosity\cite{erlumi}
will lead to a reduction of this error to $0.6\times~10^{-3}$.
\item $\Rl$ is determined by the measurement of the leptonic
 and hadronic
 cross sections. Since $\Rl$ is a ratio, the uncertainty arising
from the luminosity cancels and thus the uncertainty on $\Rl$ comes 
only from the statistical
and systematic errors in the $\eeff$ event selections which will be 
discussed in Section~\ref{sec:als}.
\end{itemize}
\begin{figure}[ht]
\begin{center} \vspace{-.5cm}
\begin{tabular}{c}
\mbox{\epsfig{file=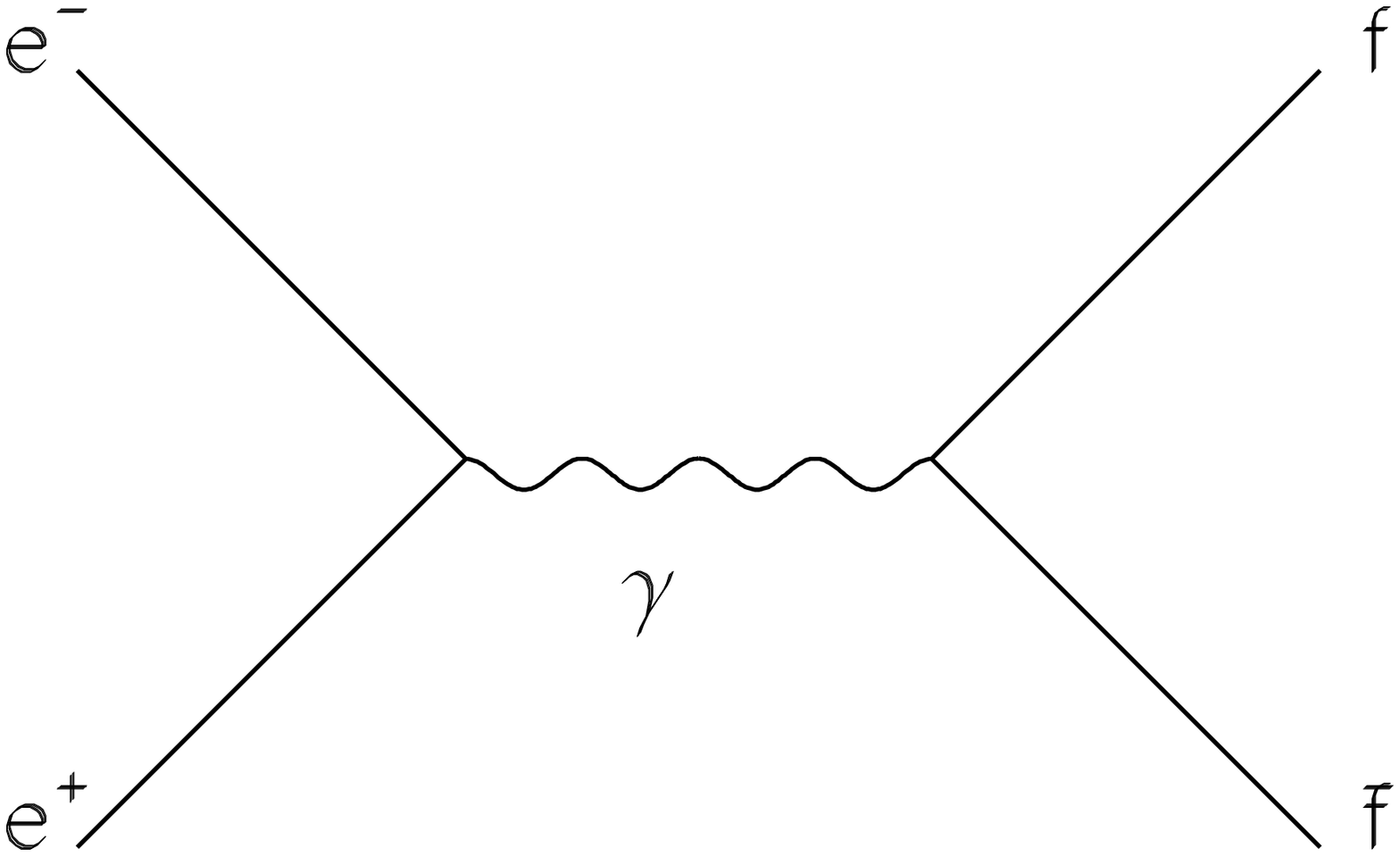,height=4cm}} \hspace{2cm}
\mbox{\epsfig{file=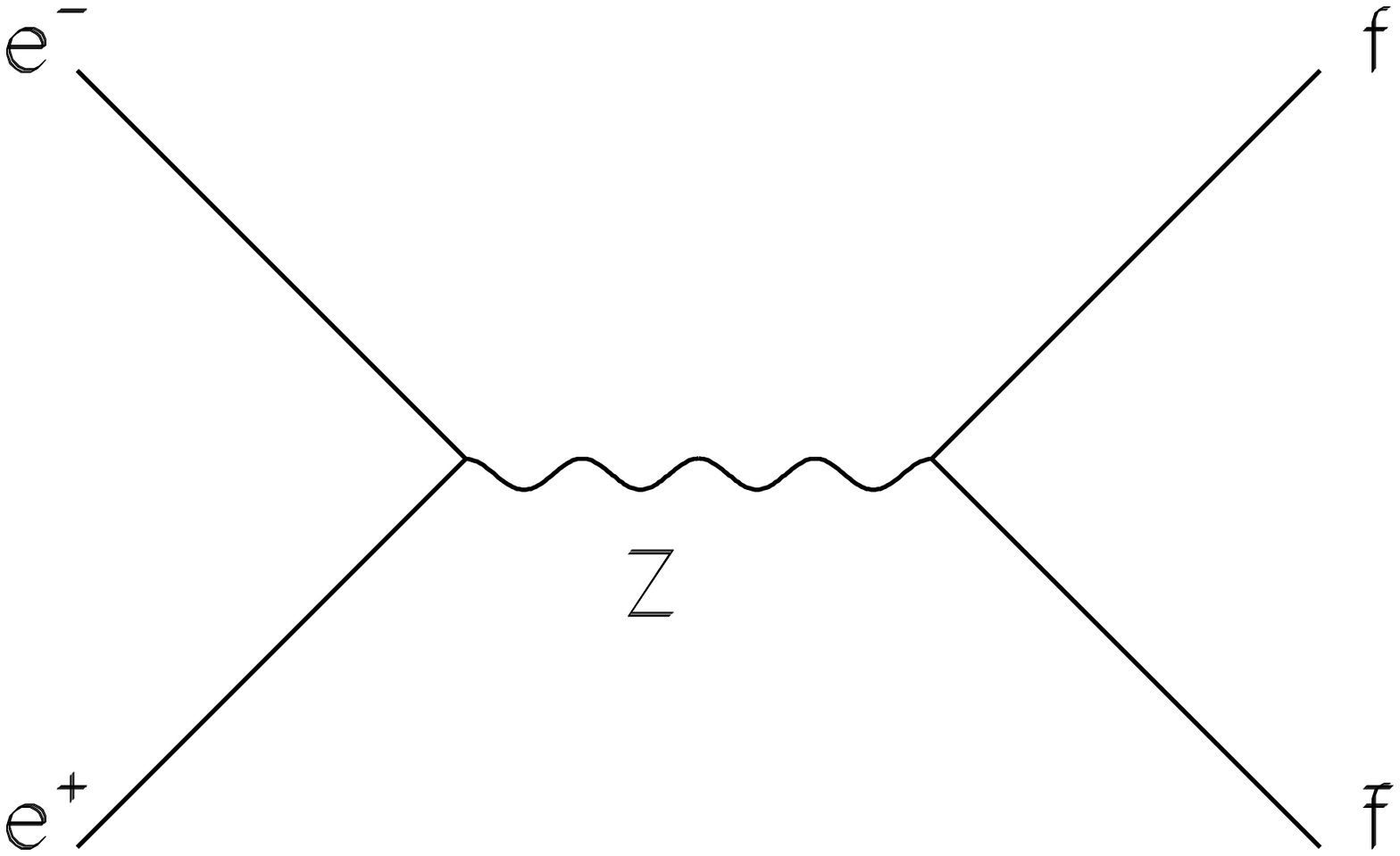,height=4cm}}
\end{tabular}
\end{center}
\vspace{-1.2cm}
\caption{\it Lowest order diagrams contributing to the $\eeff$ cross section.}
\label{fig:eeff}
\end{figure}
The lowest order diagrams involved in the process $\eeff$ are
shown in Figure \ref{fig:eeff}.
 The cross section
is the sum of the Z exchange, the $\gamma$ exchange and their
interference
\begin{equation}
\label{eq:sigtot}
\mathrm{
\sigma=\sigma_Z+\sigma_{\gamma}+\sigma_{int}
}
\end{equation}
The cross section due to the Z exchange is parametrised with a Breit-Wigner
in a model independent way using $\MZ$, $\GZ$, and the Z partial widths $\Gf$
\begin{equation}
\label{eq:sig_z}
\mathrm{\sigma_Z=\sigma^0_f \frac{s\GZt}{(s-\MZt)^2+(s\GZt/\MZt)^2}
{\hspace{0.5cm} with \hspace{0.5cm}} \sigma^0_f=\frac{12\pi\Ge\Gf}{\MZt\GZt}
}
\end{equation}
The photon contribution is taken from QED
\begin{equation}
\label{eq:sig_g}\mathrm{
\sigma_{\gamma}=\frac{4\pi\alpha^2}{3s}Q_e^2Q_f^2N_f^C
}
\end{equation}
where $\mathrm{N_F^C}$ is the number of colours for quarks and
 1 for leptons.
The interference term contains combinations of the couplings of the fermions
to the Z and their charge, which cannot be expressed in terms of the Z 
parameters used in equation~\ref{eq:sig_z}
 and are included in the $\mathrm{J_f}$ parameter
\begin{equation}\mathrm{
\sigma_{int}=\frac{4\pi\alpha^2}{3s}J_f\frac{s-\MZt}{(s-\MZt)^2+(s\GZt/\MZt)^2}
}
\end{equation}
Since the interference term is expected to be small it is usually set to the 
Standard Model value introducing a small model dependence in the fit.\\
The effect of initial state radiation is taken into account in the fitting 
procedure through its convolution with the Born cross section. 
\begin{table}
\begin{center}
\begin{tabular}{|c|c|}
\hline
Parameter & LEP average\\
\hline
$\MZ$ & $91.1867\pm0.0020$ \\
\hline
$\GZ$ & $2.4948\pm0.0025$ \\
\hline
$\sigh$ & $41.486\pm0.053$ \\ 
\hline 
$\Rl$ & $20.775\pm0.027$ \\
\hline
\end{tabular}
\end{center}
\vspace{-.5cm}
\caption{\it Average line shape parameters from the results of the four LEP
experiments.}\label{tbl:Zpar}
\end{table}
The fitted Z lineshape parameters are given in  Table~\ref{tbl:Zpar}.
These are the physical parameters, i.e they include 
all electroweak and strong radiative corrections
and QED final state corrections:
\begin{equation}
\mathrm{
\Gf=\frac{G_F M^3_Z}{6\pi\sqrt{2}}\biggl(g_{A,f}^2+g_{V,f}^2\biggr)
\biggl(1+\delta_{QED}\biggr)\biggl(1+\delta_{QCD}\biggr)
}
\end{equation}
The QCD correction $\delta_{QCD}$ is zero for leptons 
and is to a first approximation proportional to $(1+\als/\pi)$ for 
hadrons. The coupling constants $\mathrm{g_{A,f}}$ and $\mathrm{g_{V,f}}$
 absorb the electroweak corrections through $\Delta \rho$ and $\sineff$:
\begin{equation}
\mathrm{g_{A,f}}=\mathrm{\sqrt{1+\Delta \rho }\times I_3} \hspace{1cm}
\mathrm{g_{V,f}}=\mathrm{\sqrt{1+\Delta \rho }\times(I_3-2Q_f\sineff )}
\end{equation}
The dependence of $\Delta \rho$ and of $\sineff$ on $\Mt$ is quadratic,
while the dependence on the Higgs mass is only logarithmic giving less 
sensitivity to $\MH$.  These radiative corrections are detailed in 
\cite{YR95}.
\\
The measured values of the Z parameters
are confronted with the Standard  Model prediction in order to
extract the radiative corrections and therefore to determine $\als$,
$\MH$ and $\Mt$.
This fit is performed using the latest version of 
the programs ZFITTER and TOPAZ0 \cite{zf} which
include new calculations of radiative corrections:
\begin{itemize}
\item non factorisable QCD/EW corrections \cite{als_fac} resulting
in an increase of the fitted value of $\als$ of +0.001 with respect 
to old versions.
\item two loop irreducible EW corrections \cite{2loopew} 
$\mathrm{{\cal O}(\alpha^2M^2_t/M^2_W)}$ 
leading to a decrease in the Higgs mass of about 30~GeV/c$^2$. 
\item four loop QCD corrections in the $\beta$ function \cite{4loopqcd}.
\end{itemize}
 Results will be given in the next Sections.
%
%
\section{Extraction of $\als$}
\label{sec:als}
\begin{figure}[h]
\begin{center}\vspace{-.5cm}
\mbox{\epsfig{file=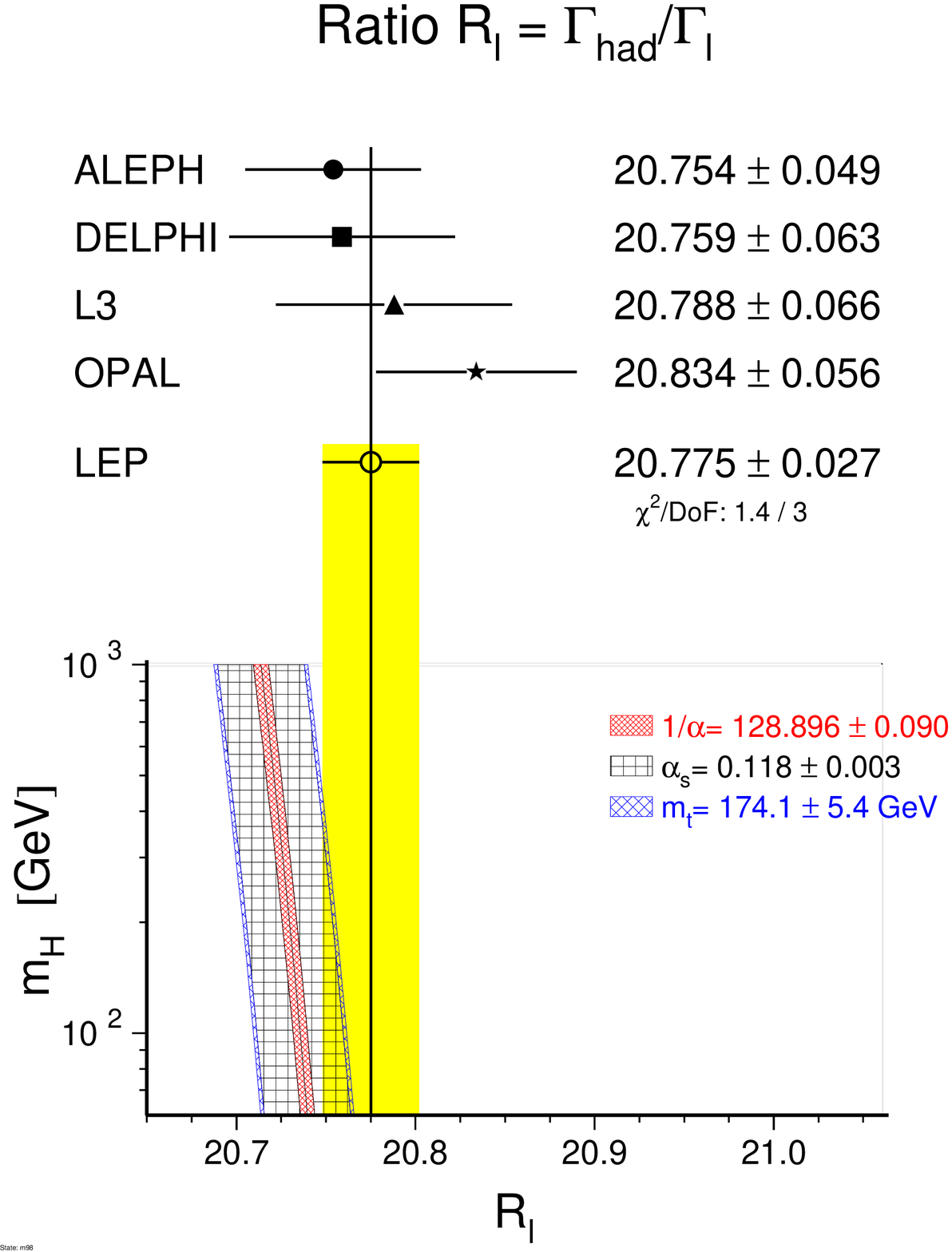,height=9.cm}}
\end{center}
\vspace{-.5cm}
\caption{\it Determination of $\Rl$ at LEP. The Standard Model prediction as a 
function of $\MH$ is also shown. The width of the Standard Model band 
corresponds to the uncertainties on $\als$, $\Mt$ and $\alem$. The total width
of the band is the linear sum of these effects.}
\label{fig:Rl_lep}
\end{figure}

\begin{figure}[h]
\begin{center}
\mbox{\epsfig{file=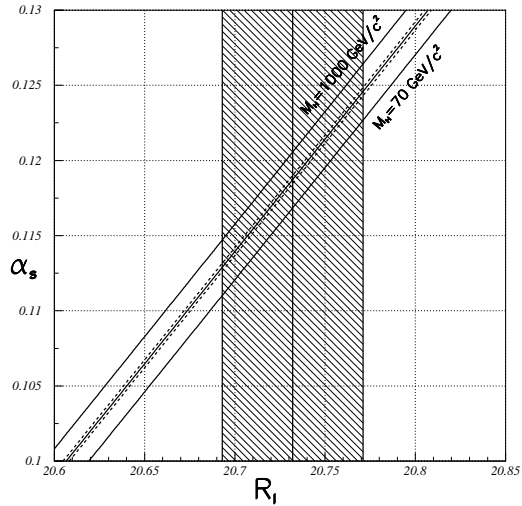,height=7.5cm}}
\end{center}
\vspace{-.5cm}
\caption{\it ALEPH determination of $\Rl$ (hatched band) and 
Standard Model expectation as a function of $\als$.
The full lines give the SM prediction for $\Mt=174.1$~GeV/$c^2$ and 
$\MH=$70, 300, and 1000~GeV/$c^2$. The doted lines correspond to a variation of
$\Mt$ of $\pm5$~GeV/$c^2$ for $\MH$ fixed to 300~GeV/$c^2$.}
\label{fig:als_Rl}
\end{figure}

The most sensitive  parameters to $\als$ are $\GZ$, $\sigh$
and  $\Rl$.
They all depend on $\als$ through $\Gh$. Their dependence with 
$\als$ is given in a first approximation by:
\begin{equation}
\Rl=\frac{\Gh}{\Gl}\propto\biggl(1+\frac{\als}{\pi}\biggr)
\end{equation}
\begin{equation}
\mathrm{\GZ=\Gh+\Ge+\Gm+\Gt+\Ginv\propto\biggl(1+0.7\frac{\als}{\pi}\biggr)}
\end{equation}
\begin{equation}
\mathrm{\sigh\propto\frac{\Ge\Gh}{\GZt}\propto\biggl(1-0.4\frac{\als}{\pi}\biggr)}
\end{equation}
Table \ref{tbl:sens_als} gives the relative experimental precision
on these parameters and the derived uncertainty on $\als$ when 
 determined  using only the corresponding parameter.
\begin{table}[h]
\begin{center}
\begin{tabular}{|c|c|c|}
\hline
       & $\Delta x/x$ & $\Delta \als$\\
\hline
$\Rl$  &  1.3 $10^{-3}$ & 0.004\\
$\GZ$  &  1.0 $10^{-3}$ & 0.004\\
$\sigh$  &  1.3 $10^{-3}$ & 0.010\\
\hline
\end{tabular}
\end{center}
\vspace{-.5cm}
\caption{\it Relative experimental precision on the parameters used in $\als$
determination and induced uncertainty on $\als$.}
\label{tbl:sens_als}
\end{table}
\begin{figure}[h]
\begin{center}\vspace{-.5cm}
\mbox{\epsfig{file=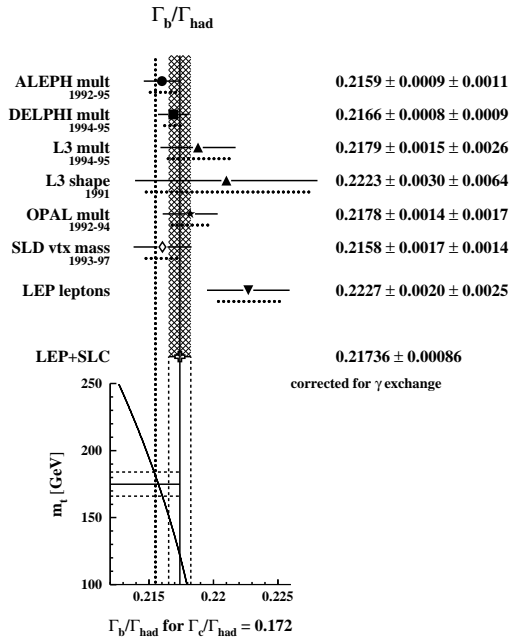,height=9.cm}}
\end{center}
\vspace{-.5cm}
\caption{\it Experimental determination of $\Rb$ at LEP and SLC. Also
shown is the Standard Model expectation as a function of $\Mt$ and the 
direct top mass measurement $\Mt=(174.1\pm5.4)$~GeV/c$^2$\cite{mtop}.}
\label{fig:Rb}
\end{figure}
$\Rl$ and $\GZ$ both allow to determine $\als$ with the 
same experimental precision\footnote{The error on $\sigh$ should decrease from
 $1.3\times10^{-3}$ to about $0.9\times10^{-3}$ with the decrease 
of the theoretical error on the 
luminosity\cite{erlumi} but $\sigh$ will still be less powerful than $\Rl$ to
determine $\als$.}
 but $\GZ$ varies rapidly with $\MH$
introducing an additional uncertainty if no constraint on $\MH$ is used
whereas the $\MH$ dependence of $\Rl$ almost cancels in the ratio of the
 hadronic to leptonic widths.
Therefore $\Rl$ is the most sensitive to $\als$ and allows to determine 
$\als$ without any assumption on the Higgs mass.\\
\\
One should note that $\Rl$, $\GZ$ and $\sigh$ all depend on $\Gb$ through
$\Gh$ and that $\Rb=\Gb/\Gh$ is fixed to its Standard Model value
in order to extract $\als$.
However, since $\Rb$ is very sensitive to new physics through
the $\mathrm{Zb\bar{b}}$ vertex corrections its value may deviate from 
Standard Model expectation. The 
experimental determination of $\Rb$
is shown in Figure \ref{fig:Rb} and is in agreement within 1.8$\sigma$
with the SM prediction for the direct top mass measurement
 $\Mt=(174.1\pm5.4)$~GeV/$c^2$\cite{mtop}. 
This value of $\Mt$ is used in the following.
\\\\
{\bf The measurement of $\Rl$}\\
As already discussed in Section~\ref{sec:Zline}, the uncertainty on
$\Rl$ is dominated by the dilepton statistical and systematic 
experimental errors. The preliminary measurements of the 4 LEP experiments
\cite{lepew98} presented in Jerusalem~97 are shown in Figure~\ref{fig:Rl_lep}.
The statistical uncertainty is 
$\Delta \Rl / \Rl (stat)\simeq (0.05(q\bar{q})\oplus 0.15 
(l^+l^-))\% $ for each LEP experiment. The systematic uncertainty 
from the hadronic channel is of the order of $0.08\%$ and an effort is being
made to reduce the systematic error from the leptonic channel.
\\
A new measurement of $\Rl$ using ALEPH data with a global dilepton
selection allows to reduce the systematic uncertainty from the leptonic
channel to $0.08\%$.
The dilepton selection has to be inclusive in order to be compared to
 the theoretical prediction. Therefore this selection
 includes also four fermion final state events of the type 
$\mathrm{\ell^+\ell^-V}$ where $\ell$ is a lepton and V is a low
invariant mass pair of fermions.
\\
First, dileptons are selected with an efficiency of $99.2\%$ inside the 
detector acceptance and background from $\gamma\gamma$, $\qq$ and 
cosmic events is reduced to the level of $0.2\%$. All the systematic 
uncertainties are at the level of few $10^{-4}$.
Then the lepton flavour separation is performed inside the dilepton 
sample so that the systematic uncertainties are anti-correlated between
 2 lepton species and that no additional uncertainty is introduced on
$\Rl$. This separation is needed in order to identify $\eeee$ events
for which the t-channel contribution has to be subtracted. This subtraction
is performed using theoretical calculation described in \cite{alistar}.
The theoretical error assigned to the subtraction is $0.08\%$ of the $\eeee$
s-channel cross section\cite{tch}.
\\
The value of $\Rl$ obtained with this new selection is 
\begin{equation}
\Rl = 20.732 \pm 0.038 \hspace{1cm}\mathrm{(ALEPH)}
\end{equation}
The relative error on $\Rl$ is reduced from $2.4\times 10^{-3}$ (1997 value)
to  $1.9\times 10^{-3}$ (this measurement).
\\\\
{\bf Determination of $\als$ with $\Rl$}\\
The dependence of $\Rl$ on $\als$ is parametrised with the latest
version of ZFITTER\cite{zf}:
\begin{equation}
\Rl=19.934\biggl(1+1.045\biggl(\frac{\als}{\pi}\biggr)+
0.94\biggl(\frac{\als}{\pi}\biggr)^2
-15\biggl(\frac{\als}{\pi}\biggr)^3\biggr)
\end{equation}
In this parametrisation $\MH=300$~GeV/c$^2$ and $\Mt=174.1$~GeV/c$^2$ are used.
The small dependence with the Higgs and the top masses is also parametrised 
with ZFITTER
\begin{equation}
\Rl\propto 
\biggl(1-2.2\times10^{-4}ln\biggl(\frac{{\MH}}{\MZ}\biggr)^2\biggr)\times
\biggl(1-4.1\times10^{-4}\biggl(\frac{\Mt}{\MZ}\biggr)^2\biggr)
\end{equation}
The values of $\Rl$ obtained with this parametrisation agree with ZFITTER
prediction at the level of few $10^{-5}$ for any value of $\als$ in
[0.100,0.130], $\Mt$ in [150,200]~GeV/c$^2$ and $\MH$ in [60,1000]~GeV/c$^2$.
Figure \ref{fig:als_Rl} shows the new ALEPH determination of $\Rl$ and 
the Standard Model
expectation as a function of $\als$. The variations arising from
$\MH$ and $\Mt$ are also shown.
The value obtained for $\als$ is
\begin{equation}
\alsmz=0.119\pm0.006_{exp}\pm0.002_{QCD}\pm0.002_{\MH} 
\hspace{2mm} \mathrm{(ALEPH)}
\label{eq:fittorl_al}
\end{equation}
where the first error comes from the experimental uncertainty on $\Rl$,
the second error covers missing higher-order corrections and 
uncertainties in the interplay of electroweak and QCD corrections\cite{als_er}
 and the last error
is obtained by varying the Higgs mass from 60 to 1000~GeV/$c^2$.
\\ 
The preliminary
 LEP combined value or $\Rl$ (Jerusalem 97, see Ref.\cite{lepew98}) 
is shown in Figure  \ref{fig:Rl_lep} and leads to 
\begin{equation}\mathrm{
\alsmz=0.125\pm0.004_{exp}\pm0.002_{QCD}\pm0.002_{\MH}
}\label{eq:fittorl}
\end{equation}
In an overall fit using all the information from LEP1 and SLC data
(cross sections and asymmetries) the value of $\als$ and of $\MH$ are
simultaneously constrained (see Section~\ref{sec:Mhiggs}), the fitted value
of $\als$ is
\begin{equation}\mathrm{
\alsmz=0.120\pm0.003_{exp}\pm0.002_{QCD}
}
\end{equation}
The experimental error is reduced from 0.004 (in equation~\ref{eq:fittorl})
to 0.003
because $\GZ$ and $\sigh$ are used in this last fit,
and the 0.002 error from the Higgs mass disappears since $\MH$ is also
constrained. This value of $\als$ is lower than  in the fit to $\Rl$ 
 alone (equation~\ref{eq:fittorl})
because $\GZ$ prefers lower values of $\als$ and the low fitted value
of $\MH$ (66~GeV/c$^2$) brings $\als$ down by $\sim$0.002 (in the fit to $\Rl$ the
value $\MH=300$~GeV/c$^2$ was used).
This value is in good agreement with the world average 
$\alsmz=0.118\pm0.003$\cite{als_wa} and of comparable precision.
%
%
\section{Constraint on the SM Higgs mass}
\label{sec:Mhiggs}
\begin{figure}[h]
 \begin{minipage}{.48\linewidth}
        \begin{center}
         \mbox{\hspace{-1.5cm}\epsfig{file=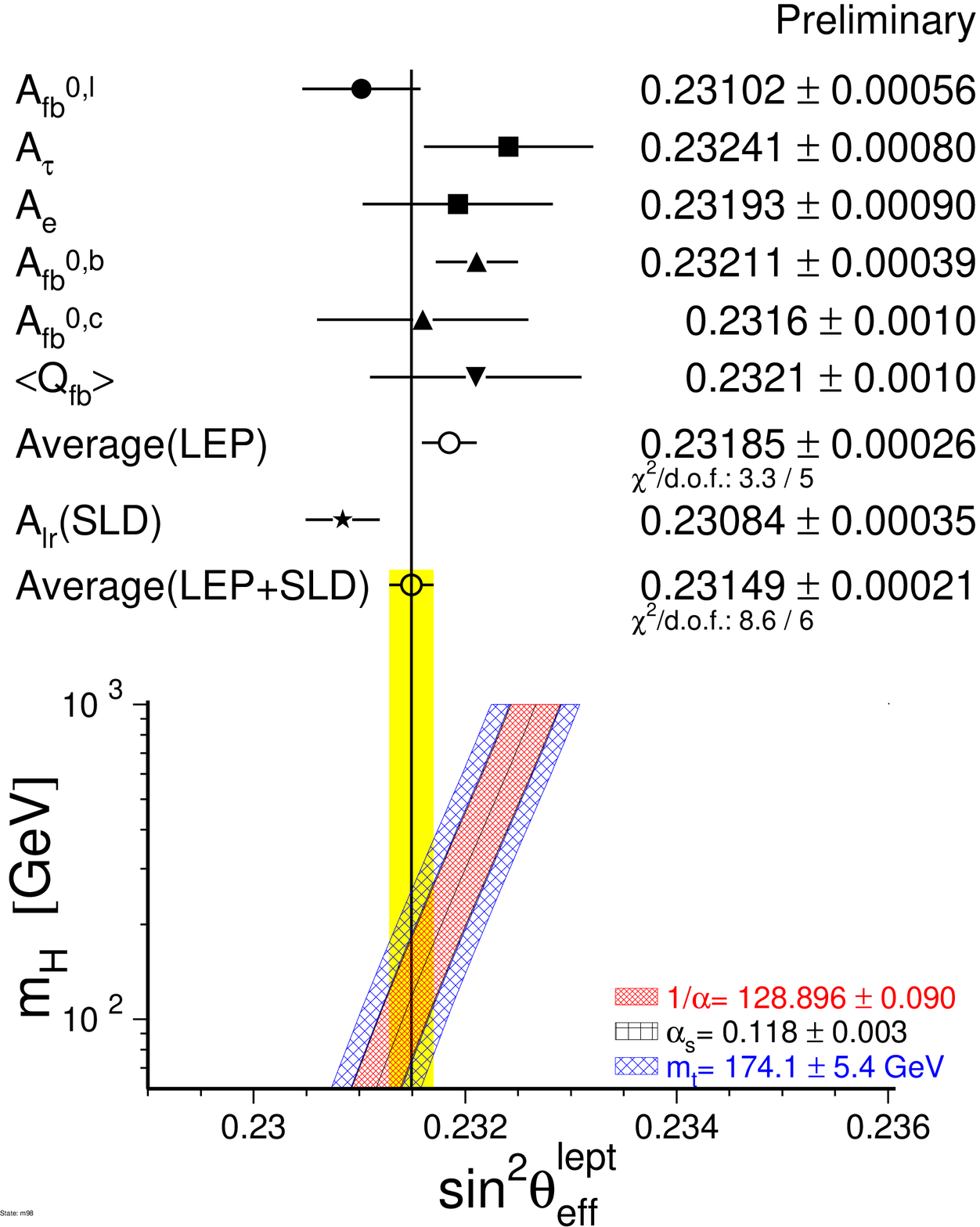,height=9.cm}}
        \end{center}
	\vspace{-.5cm}
        \caption{\it Determinations of $\sineff$ from the asymmetries.
The prediction from the Standard Model as a function of $\MH$ is also shown.
The width of the Standard Model band is due to the uncertainties in
$\alem$, $\alsmz$ and $\Mt$.}
\label{fig:sineff}
       \end{minipage}\hfill
       \begin{minipage}{.48\linewidth}
        \begin{center}
         \mbox{\hspace{-.5cm}\epsfig{file=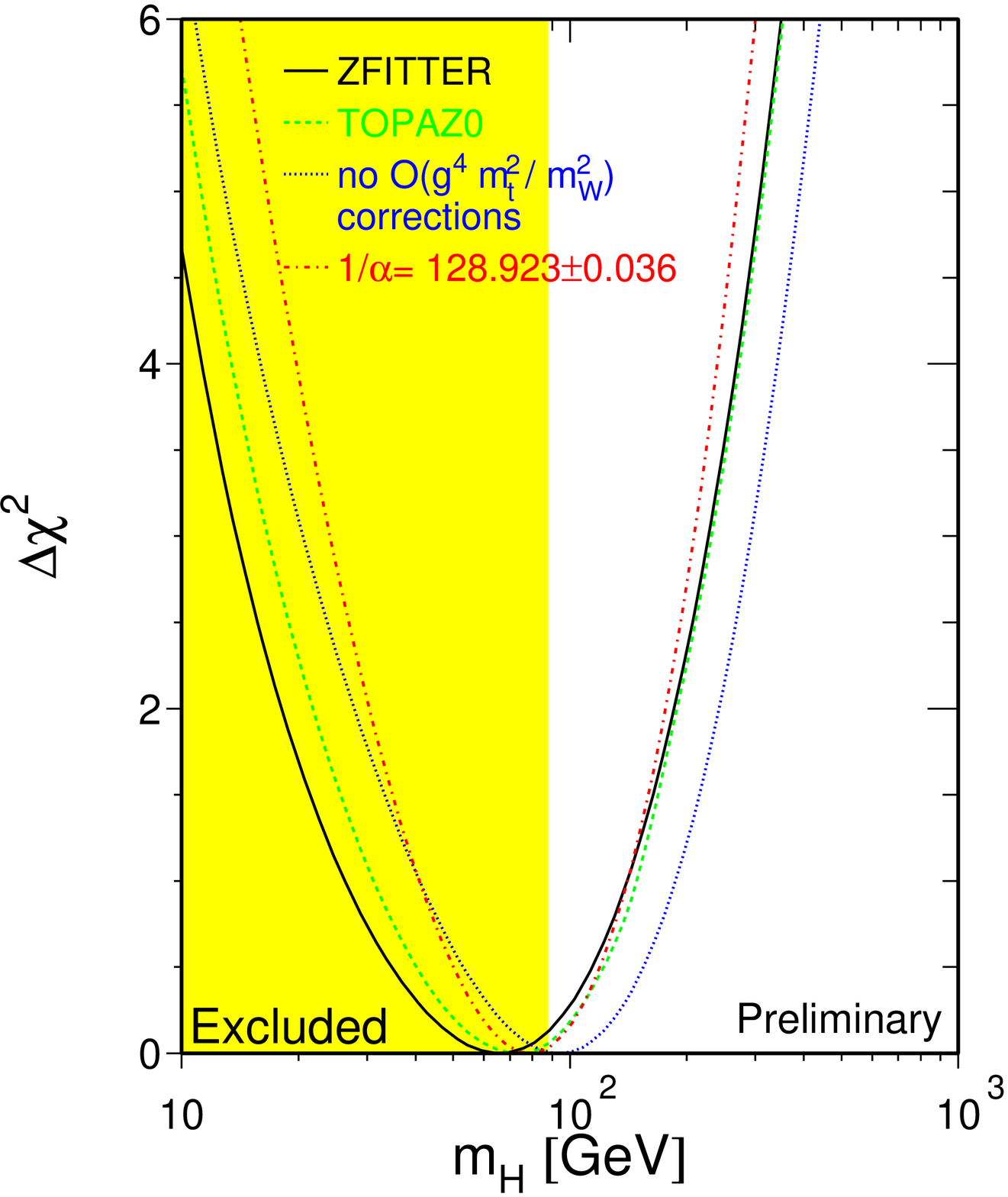,height=9.cm}}
        \end{center}
	\vspace{-.5cm}
        \caption{\it $\Delta \chi^2=\chi^2-\chi^2_{min}$ vs. $\MH$ curve.
The line is the result of the fit using all data (last column of Table 
\ref{tbl:fit}). The vertical band shows the 95$\%$ CL exclusion limit on $\MH$
from direct searches \cite{MH_direct}.}\label{fig:mh}
       \end{minipage}
\end{figure}

{\bf The determination of $\sineff$}\\
At the Z resonance, the most sensitive parameter to the Higgs mass
is $\sineff$ which is determined through the measurement of the asymmetries
\begin{itemize}
\item the lepton, b and c quarks Forward-Backward asymmetries
$\mathrm{A^{0,f}_{FB}=\frac{3}{4}{\cal A}_e{\cal A}_f}$
determined at LEP1 from $\eeff$ angular distributions
\begin{equation}\mathrm{
\frac{d\sigma}{dcos\theta}\propto 1+cos^2\theta+\frac{8}{3}A_{FB}cos\theta}
\end{equation}
\item The $\tau$ polarisation (LEP1) 
 \begin{equation}
\mathrm{ {\cal P}_{\tau}(cos\theta)=
-\frac{{\cal A}_{\tau}(1+cos^2\theta)+2{\cal A}_{e}cos\theta}
{1+cos^2\theta+2{\cal A}_{e}{\cal A}_{\tau}cos\theta}}
\end{equation}
\item the Left-Right asymmetries measured at SLC with polarised electron beam
\begin{equation}
\mathrm{A_{LR}={\cal A}_e}
\end{equation}
\end{itemize}
where $\mathrm{{\cal A}_f=2(g_{V,f}/g_{A,f})/(1+(g_{V,f}/g_{A,f})^2)}$.
All these quantities are expressed in terms of the effective mixing angles of
 leptons, $\sineff$. The derived values of $\sineff$ from the 
different asymmetry measurements \cite{lepew98} are compared in 
Figure~\ref{fig:sineff}.
SLD and LEP average differ by 2$\sigma$ and SLD data prefer
 lower values of the Higgs mass. Since all LEP data
have been analysed new measurement can only come from SLC with increased
statistics.\\\\
{\bf The global fit}\\
In the global fit the electroweak
measurements are compared to Standard Model predictions
in order to determine simultaneously the values of $\alsmz$, $\Mt$ and $\MH$.
The parameters 
$\mathrm{G_F=(1.16639\pm0.00002)\times 10^{-5}}$~GeV$^{-2}$ from muon decay
and $\alem^{-1}=128.896\pm0.090$ from \cite{al_qed95}
are also used as input in the fit.
This global fit is performed with three data sets, results are given  
in Table \ref{tbl:fit}:
\begin{enumerate}
\item A first fit to
 LEP data alone including the measurement of the W mass at LEP 2
is made (first column of Table~\ref{tbl:fit}).
This fit shows that LEP data prefer a light top and a light Higgs.
\item Then LEP1 and SLD data and the measurement of $\sinW$ from $\nu$N
experiments are used to determine 
top quark and W masses indirectly (second column of Table~\ref{tbl:fit}). 
The good agreement with the direct 
measurements $\Mt=174.1\pm5.4$~GeV/c$^2$\cite{mtop} and $\MW=80.430\pm0.080$~GeV/c$^2$
\cite{mwcomb} provides a test of the Standard Model.
\item All electroweak measurements including the direct $\Mt$ and 
$\mathrm{M_W}$ measurement are used to obtain the best constraint on $\MH$
(last column of Table~\ref{tbl:fit}).
Figure \ref{fig:mh} shows the result of this fit using the latest version of
ZFITTER and TOPAZ0. 
Also shown is the curve obtained when new higher-order corrections are 
neglected (`no $\mathrm{{\cal O}(g^4M^2_t/M^2_W)}$' curve). 
These corrections lead
to a decrease of the Higgs mass of $\sim30$~GeV/c$^2$.
The upper limit on $\MH$ is 215~GeV/c$^2$ at 95$\%$ CL.
The uncertainty on $\alem$ causes an error of 0.2 on Log($\MH$).
The fit is also performed with the new evaluation of $\alem$\cite{al_qed97}.
With this value the fitted error on Log($\MH$) is reduced by $30\%$
(from 0.33 to 0.25).
\end{enumerate}
Since $\MH$ is mainly constrained by the measurement of $\sineff$, new
data from SLC will improve this constraint.
Moreover, radiative
corrections are shared between the top quark and the Higgs, therefore the 
reduction of the error on the direct top mass measurement will improve the 
constraint on the Higgs mass. The W mass is also sensitive to the Higgs mass
and its measurement with an error of 30~MeV at LEP2 will also improve the
constraint on $\MH$.
\begin{table}
\begin{center}
\begin{tabular}{|c||c|c|c|}
\multicolumn{4}{c}{ }\\
\hline
  & { LEP including  }      & { all data except  }    &  { all data} \\
  & { LEP2 $\mathrm{M_W}$}       & { $\mathrm{M_t}$ and $\mathrm{M_W}$   }   &           \\
\hline
\hline
$\mathrm{M_t}$\hfill[GeV] & { $157^{+12}_{-10}$} & { $161^{+9}_{-8}$ } & { $171.1\pm5.1$  }   \\
$\mathrm{M_H}$\hfill[GeV] & { $56^{+101}_{-31}$} & { $33^{+45}_{-17}$ } & { $66^{+74}_{-39}$}\\
$\mathrm{Log(M_H/GeV)}$  & { $1.75^{+0.45}_{-0.35}$} & { $1.53^{+0.37}_{-0.29}$ }
                                                     & { $1.82^{+0.33}_{-0.40}$}\\
$\mathrm{\alpha_s(M_Z^2)}$          & { $0.122 \pm 0.003$} & { $0.121\pm0.003$ }  & { $0.120 \pm 0.003$} \\
\hline
$\mathrm{\chi^2}$/d.o.f.{} & { $ 6/9$ }           & { $14/12$   }       & { $17/15$  }         \\
\hline
\hline
$\mathrm{sin^2\theta_W^{eff}}$         & { $0.23183\pm0.00025$}
                    & { $0.23145\pm0.00021$}
                    & { $0.23146\pm0.00022$} \\
$\mathrm{sin^2\theta_W}$             & { $0.2245\pm0.0008$}
                    & { $0.2235\pm0.0008$}
                    & { $0.2230\pm0.0005$  }   \\
$\mathrm{M_W}$\hfill[GeV]   & { $80.302\pm0.040$}
                    & { $80.351\pm0.040$}
                    & { $80.380\pm0.027$  }      \\
\hline
    \end{tabular}
  \end{center}
\caption{\it Results of the electroweak fits. See text for details.}
\label{tbl:fit}
\end{table}
%
%
\section{Conclusion}
\label{sec:concl}
The large sample of data accumulated at LEP1 allows the determination
of $\als$ through radiative corrections, 
$\als=0.120\pm0.003_{exp}\pm0.002_{QCD}$
with a precision comparable to
the world average (in which LEP1 data have not been included). Since all
LEP1 data have been analysed, this determination may only slightly improve
through the reduced error on $\sigh$\cite{erlumi}.
\\
The global fit gives for the Higgs mass
\begin{equation}
\MH=66^{+74}_{-39} \hspace{.3cm} \mathrm{GeV}/c^2
\end{equation}
There is still room for improvement in this constraint through better
top and W mass direct measurements, 
new data from SLC and the reduction of the error on $\alem$.\\
A low value of the Higgs mass is preferred by the Z resonance data: 
it is still possible to find the Higgs at LEP2~!
\section*{Acknowledgements}
I would like to thank the LEP Electroweak Working Group who made the 
fit results available 
and in particular C. Paus who provided me some plots. Many thanks to the 
Quarks'98 organisers for the very pleasant atmosphere at the conference.
It is a pleasure to thank J. Lefran\c{c}ois and I. Videau for careful
 reading and comments on this manuscript.

\end{document}